# Low Carrier Density Epitaxial Graphene Devices On SiC

*Yanfei Yang*, Lung-I Huang, Yasuhiro Fukuyama, Fan-Hung Liu, Mariano A. Real, Paola Barbara, Chi-Te Liang, David B. Newell, and Randolph E. Elmquist*

Y. Yang, L.-I. Huang, D. B. Newell, R. E. Elmquist
National Institute of Standards and Technology (NIST),
Gaithersburg, MD 20899-8171
E-mail: yanfei.yang@nist.gov

Y. Yang, P. Barbara
Department of Physics
Georgetown University
Washington, DC 20057-1228

L.-I. Huang, C.-T. Liang
Department of Physics
National Taiwan University
Taipei, 10617, Taiwan

Y. Fukuyama
National Metrology Institute of Japan (NMIJ/AIST)
Tsukuba, Ibaraki 305-8563, Japan

F.-H. Liu, C.-T. Liang
Graduate Institute of Applied Physics
National Taiwan University
Taipei, 10617, Taiwan

M. A. Real
Instituto Nacional de Tecnologia Industrial
San Martin, Buenos Aires, B1650WAB, Argentina

## Introduction

Graphene, the first genuinely two-dimensional material,[1] is a single layer of carbon forming a simple hexagonal lattice. Remarkable electrical properties such as high mobility and an anomalous quantum Hall effect (QHE) were demonstrated in exfoliated graphene flakes and attracted enormous interest for many practical applications.[2-5] However, mechanical exfoliation of graphite cannot provide graphene suitable for commercial wafer-size electronics or precise resistance metrology due to the size limitation of the exfoliated graphene flakes (usually up to tens of microns).

When SiC substrates are annealed at temperatures above 1000 °C in ultra-high vacuum or an inert gas atmosphere, carbon remains on the SiC surface after Si sublimation and rearranges to form graphene layers. This epitaxial graphene (EG) is ready for large-scale device fabrication without transfer to another insulating substrate. Graphene grown on the silicon-terminated face (Si-face) of hexagonal SiC wafers can form large domains due to registry with the azimuthal orientation of the SiC crystal. On the Si-face, EG also has more controllable growth kinetics, compared to the graphene grown on the opposite (carbon) face.

Recently, growth of homogeneous monolayer EG on the Si-face of SiC has been improved by optimizing the annealing temperature and background gas conditions.[6-10] QHE plateaus have been observed in various magneto-transport measurements and the robust quantized Hall resistance (QHR) plateau with filling factor $\nu = 4(n + 1/2) = 2$ has been shown to be equivalent to that of conventional 2D electron systems based on semiconductor heterostructures at low temperature (<4 K), which are the basis of present-day electrical resistance metrology.[11-18] EG devices can operate at currents and temperatures that are considerably higher, however, for use at reasonable magnetic field levels, QHR devices must have carrier densities $n \leq 10^{12}$ $cm^{-2}$. The accepted best practices for precise QHR measurements also specify near-zero channel longitudinal resistivity with low, ohmic contact resistance values.[19] So far, only a few EG devices exist that display all of these qualities.

Strong electron (n-type) doping is imparted to Si-face EG by a buffer layer that is covalently bonded to the SiC substrate. Fermi energies determined by *in situ* angle-resolved photo-emission spectroscopy (ARPES) confirm that an intrinsic n-type doping near $10^{13}$ $cm^{-2}$ exists in as-grown EG.[16,20,21] Various techniques such as gating and chemical doping have been developed to compensate intrinsic doping levels.[11,12,15,16,22]

Here we report high mobility and low carrier density in un-gated EG devices that are produced with high yield when a metal protective layer is deposited directly on as-grown EG. This layer prevents EG contact with most organic residues (see **Figure 1**). When diluted aqua regia (DAR) is used as the final etching agent in this process, most of our Hall bar devices based on Si-face EG have carrier densities in the range of $3 \times 10^{10}$ cm$^{-2}$ to $3 \times 10^{11}$ cm$^{-2}$, much lower than those obtained by conventional lithography. Well-defined $\nu = 2$ plateaus are observed in moderate magnetic fields, and the doping level can be controlled by heat treatment.

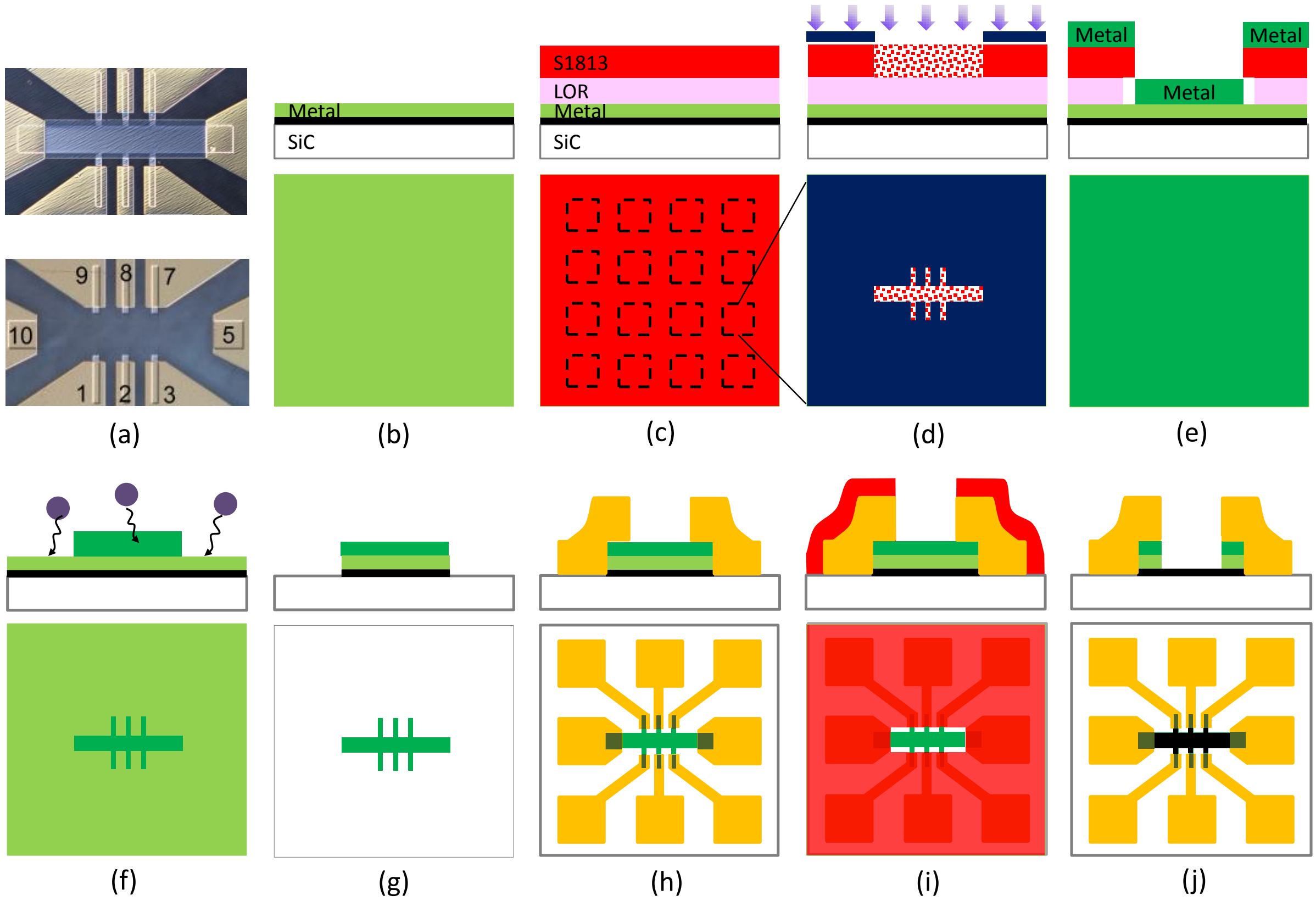

**Figure 1.** Device images and fabrication process (a) Optical images (upper) with metal protective layer covering the Hall bar area (600 μm x 100 μm) of a device and (lower) after removing the metal layer. Six 20 μm wide voltage probes (1,2,3,7,8,9) are located in the central 200 μm wide region of the Hall bar, with 100 μm longitudinal spacing. Source and drain are labeled 5 and 10. Processing steps, consisting of (b) Metal is first evaporated on as-grown graphene. (c) Spin-coating of bi-layer photoresist. Each dashed square indicates a single device area. (d) UV-exposure (with enlarged single device area shown). (e) Undercut profile is formed to assist the lift-off after evaporation of a second metal layer (green). (f) Ar reactive-ion-etching is applied. (g) Metal-protected EG Hall bar after RIE. (h) Protected graphene Hall bar with contact electrodes. (i) An etching window is opened in Hall bar area. (j) Metal over Hall bar area is etched by DAR, followed by optional stripping of photoresist using acetone.

Five samples diced from semi-insulating 4H- or 6H-SiC(0001) wafers were annealed at 1900 °C or 1950 °C in an Ar background at 101 kPa - 105 kPa using a controlled Si sublimation process (see supporting information).[8] Raman microscopy shows that homogeneous graphene covers 95 % of the central sample area for samples prepared by the same methods and under similar conditions. In the critical new processing step, a metal bilayer (5 nm Pd + 10 nm Au, for samples S1 – S3) or a single 30-nm-thick Au layer (for S4 and S5) was deposited directly on the as-grown EG surface after a 10 minutes, 115 °C dehydration baking (Figure 1b). This protective metal layer prevented the intermediate processing steps from contaminating the graphene and was removed from the Hall bars using diluted aqua regia (DAR) in the final step of fabrication (Figure 1j).

**Transport Measurements**

**Table 1.** Growth conditions and transport characteristics for five samples approximately one month after fabrication, from low-current ac measurements recorded with lock-in amplifiers; Samples S1, S2 and S3 were diced from the same 6H-SiC wafer. Samples S1 and S2 were annealed and processed together, and thus are grouped together. The carrier density and mobility were measured at 4.0 K to 4.5 K for samples S1 through S4, and at 1.6 K for sample S5. Standard deviations of the carrier densities for each sample group are low, ranging from $0.5 \times 10^{11}$ cm$^{-2}$ to $0.8 \times 10^{11}$ cm$^{-2}$.

| Device name | Wafer type | Growth Conditions (°C/min) | Electron density $n$ (cm$^{-2}$) | Mobility $\mu$ (cm$^2$V$^{-1}$s$^{-1}$) | $R_s$ (kΩ) $B$=0 |
|---|---|---|---|---|---|
| S1D1 | 6H-SiC | 1900/30 | $1.3 \times 10^{11}$ | 11000 | 4.4 |
| S1D2 | | | $1.4 \times 10^{11}$ | 6500 | 7.1 |
| S1D3 | | | $5.1 \times 10^{10}$ | 10200 | 11.9 |
| S2D1 | 6H-SiC | | $4.2 \times 10^{10}$ | 7400 | 20.2 |
| S3D1 | 6H-SiC | 1900/30 | $-5.4 \times 10^{10}$ | 7900 | 14.6 |
| S3D2 | | | $5.1 \times 10^{10}$ | 10800 | 11.3 |
| S3D3 | | | $4.0 \times 10^{10}$ | 8600 | 18.2 |
| S3D4 | | | $7.0 \times 10^{10}$ | 4400 | 20.6 |
| S3D5 | | | $6.4 \times 10^{10}$ | 8500 | 11.4 |
| S4D1 | 4H-SiC | 1900/18 | $2.0 \times 10^{11}$ | 3600 | 4.6 |
| S4D2 | | | $1.9 \times 10^{11}$ | 6000 | 5.5 |
| S4D3 | | | $3.1 \times 10^{11}$ | 2600 | 7.8 |
| S4D4 | | | $2.4 \times 10^{11}$ | 5200 | 5.1 |
| S4D5 | | | $3.2 \times 10^{11}$ | 2100 | 9.0 |
| S5D1 | 4H-SiC | 1950/15 | $8.0 \times 10^{11}$ | 5000 | 3.1 |
| S5D2 | | | $7.4 \times 10^{11}$ | 4400 | 3.8 |
| S5D3 | | | $8.2 \times 10^{11}$ | 4000 | 3.8 |
| S5D4 | | | $8.2 \times 10^{11}$ | 3300 | 4.6 |
| S5D5 | | | $6.3 \times 10^{11}$ | 4400 | 4.5 |

Each sample has 16 similar Hall-bar devices (Figure 1c). Devices having eight ohmic contacts and room-temperature sheet resistance $R_S$ lower than 21 kΩ (see Table 1) were selected for low-temperature magneto-transport characterization. All cryogenic measurements were made in a liquid-helium cryostat using a 9 T superconducting magnet. The device carrier density and mobility values determined at $T$ < 4.5 K using low-field magneto-transport are listed in Table 1, as well as the wafer properties, the maximum annealing temperature and the annealing time at that temperature. Devices on four of the samples showed homogeneous carrier densities $n$ that are lower than 3.2 x $10^{11}$ $cm^{-2}$.

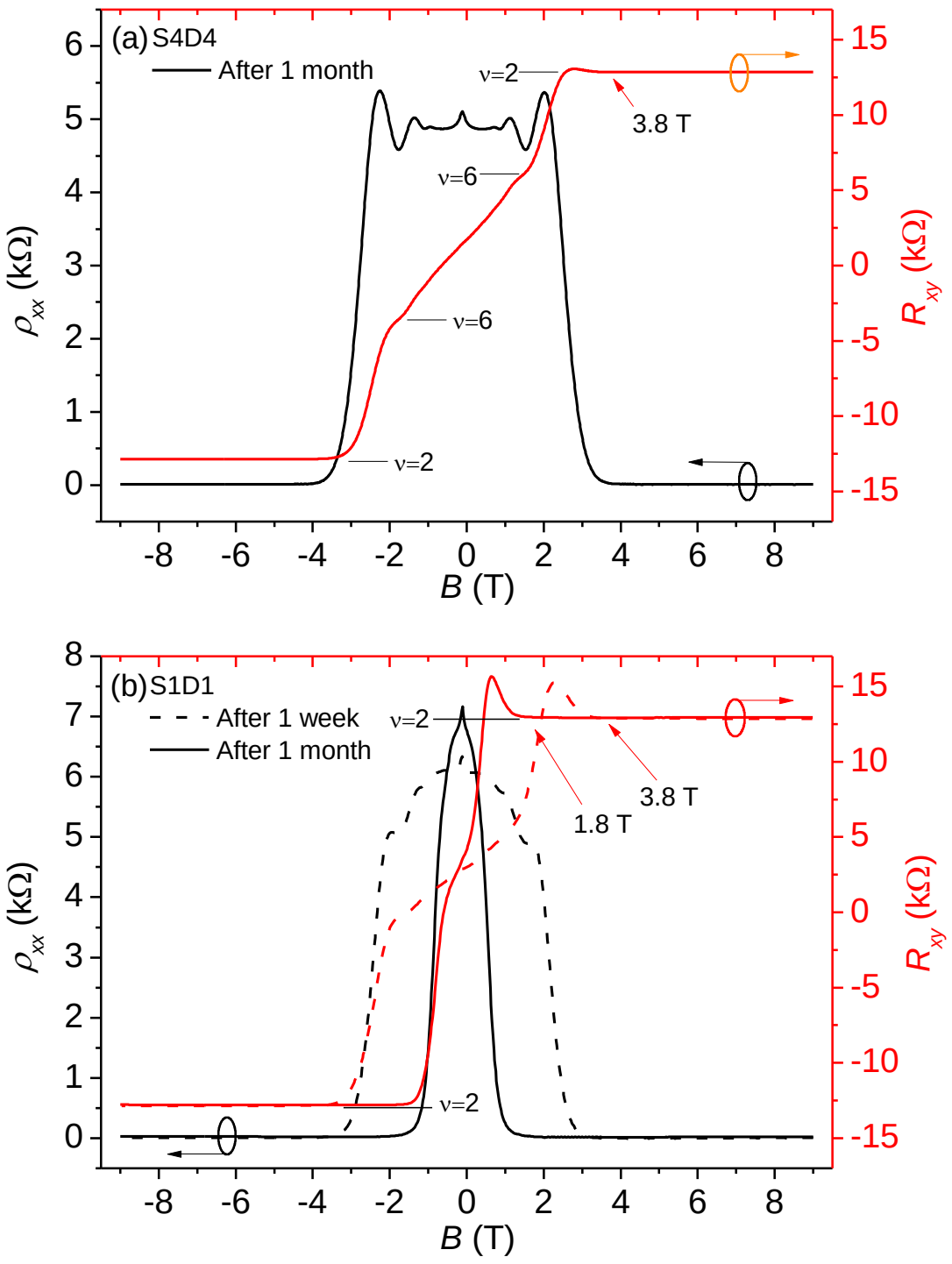

**Figure 2.** (a) Longitudinal resistivity (black) and Hall resistance (red) of device S4D4 at 4.3 K, one month after fabrication. (b) Similar measurements for device S1D1 at 4.3 K measured one week (dashed lines) and one month (solid lines) after the fabrication, respectively.

We focus first on magneto-transport measurements of device S4D4. Figure 2a shows that the carrier density and mobility at 4.3 K are $n = 2.4 \times 10^{11}$ $cm^{-2}$ and $\mu = 5200$ $cm^2V^{-1}s^{-1}$, respectively. Prominent Shubnikov-de Hass (SdH) oscillations are present with minima at the $\nu$=6 and $\nu$=10 filling factors, indicating good homogeneity along the device channel. Precise dc measurements at 1.6 K using a cryogenic current comparator bridge at a source-drain current of 40 μA show that the device is nearly fully quantized with a longitudinal resistivity of $\rho_{xx}$ = 0.05 Ω ± 0.01 Ω and a QHR at the $\nu$=2 plateau of $R_{xy} = (\frac{h}{2e^2}) \times (1 - 4 \times 10^{-6} \pm 2 \times 10^{-6})$.[23] Very nearly the same QHR characteristics were measured for both magnetic field directions. Contact resistance values in the quantized Hall plateau region were between 1 Ω and 30 Ω, with six of eight contacts showing ohmic resistance below 4 Ω.

A significant drop in n-type carrier density over time was observed during the first month after fabrication in devices S1D1 and S1D2. These samples were stored under ambient laboratory conditions (23 °C, 40 % relative humidity) when not in use. Figure 2b shows two sets of data at $T$ = 4.3 K with $\rho_{xx}$ and $R_{xy}$ measured in S1D1 using lock-in amplifiers. In the data recorded one week after the fabrication (dashed lines), a well-developed $\nu$=2 plateau is present for $B \geq 3.8$ T and the longitudinal resistivity is below 10 Ω in the high magnetic field range. In the low magnetic field region, SdH oscillations are weak. These characteristics may indicate an inhomogeneous carrier concentration along the Hall bar. The solid lines in Figure 2b were measured for device S1D1 one month after device fabrication. The $\rho_{xx}$ profile is compressed and the SdH oscillations are completely absent. We observe the onset of the $\nu$=2 Hall plateau at low fields, with $n = 1.3 \times 10^{11}$ cm$^{-2}$ and $\mu = 11\,000$ cm$^2$V$^{-1}$s$^{-1}$. The minimum longitudinal resistivity at high field has approximately doubled to 20 Ω.

Over a similar one-month time interval very low carrier concentrations were maintained on sample S3. Some devices that were slightly p-doped soon after fabrication became slightly n-doped. Only device S3D1 remained p-doped after one month in ambient conditions (see Table 1). In all of the devices on sample S3 and some devices on S1 and S2 the Fermi energy is quite close to the Dirac point. Regions dominated by electrons or holes (as puddles) are known to exist in this regime for pristine exfoliated graphene. Although readily apparent $\nu$=2 Hall plateaus are observed at very low magnetic fields in our devices on sample S3, the transport results show that the low-density devices are not well-quantized at any magnetic field level, with minimum $\rho_{xx} > 300\ \Omega$ and $R_{xy} < h/2e^2$. The effect of very low carrier concentration on transport properties in graphene close to the Dirac point is not the main issue of this paper and will be reported elsewhere.

**Post-processing Results**

Our post-processing studies show that the low carrier density results from the doping of molecules deposited by aqua regia, while the variation in the carrier density with time may be due to desorption of the doping molecules and adsorption of other molecules from the air. Atomic force microscope (AFM) images taken after fabrication show some surface roughness at a small length scales that was not present prior to device fabrication, suggesting that the EG surface may be covered with foreign matter (see Figure S1).

We subjected sample S4 to heat-treating (HT) in Ar gas (99.999 %) at relatively low temperatures, alternating with 1 s immersion in DAR (followed immediately by rinsing in deionized water). After each step, transport measurements were made. Table 2 shows the results of magneto-transport measurements on S4D4 at temperatures near 4.5 K. The n-type doping levels were raised by more than an order of magnitude when the sample was heat-treated at temperatures of 250 °C and 175 °C, and very low n-type carrier concentrations were restored by dipping the sample in DAR. When the sample was stored for six days in ambient air after HT at 175 °C, the n-type carrier concentration decreased by 63 %, but no change was observed when the sample was stored for four days in He gas at cryogenic temperatures following the HT at 250 °C. The device magneto-transport characteristics were not obviously degraded by DAR/HT cycling, and extremely well-developed QHR plateaus were observed after the final processing step shown in Table 2.

**Table 2.** Results of post-processing of device S4D4, measured at 4.54 K ± 0.04 K. DAR results in strongly reduced n-type doping, while heat treatment increases the n-type carrier concentration.

| **Device S4D4 Process, *T* (time)** | **Electron density** **$n$ (cm$^{-2}$)** | **Mobility** **$\mu$ (cm$^2$V$^{-1}$s$^{-1}$)** | **$R_s$ (kΩ)** **$B$=0** |
|---|---|---|---|
| Initial fabrication | $2.4 \times 10^{11}$ | 5200 | 5.1 |
| HT, 250 °C (2.5 h) | $2.9 \times 10^{12}$ | 2030 | 1.0 |
| Helium, <30 K (100 h) | $2.9 \times 10^{12}$ | 2030 | 1.0 |
| DAR, 20 °C (1 s) | $4.0 \times 10^{10}$ | 16800 | 10.3 |
| HT, 175 °C (2.5 h) | $1.8 \times 10^{12}$ | 2800 | 1.3 |
| Ambient air (140 h) | $6.8 \times 10^{11}$ | 6100 | 1.5 |
| DAR, 20 °C (1 s) | $-5.0 \times 10^{10}$ | 10200 | 12.1 |
| HT 100 °C (0.5 h.) | $2.9 \times 10^{11}$ | 5160 | 4.2 |

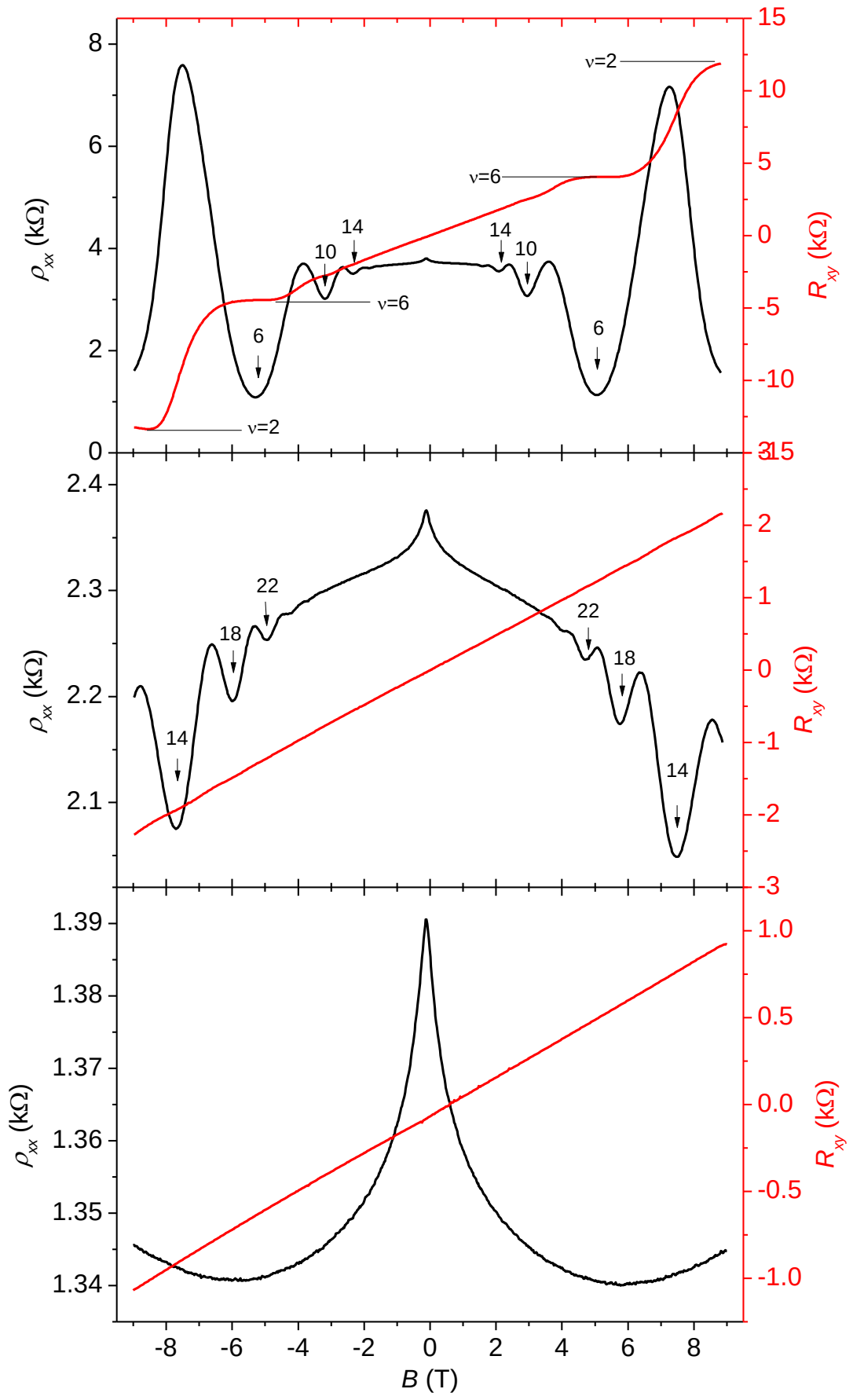

**Figure 3.** Magneto-transport results ($\rho_{xx}$ and $R_{xy}$) for device S5D2 at 1.5 K (note changes in scale for the vertical axes): (a) as fabricated with the 30-nm-thick Au layer etched by aqua regia; (b) after spin-coating using PMMA and curing; (c) after removal of PMMA. Landau filling factors are used to identify some features.

We conclude with results of processing sample S5 with Polymethyl methacrylate (PMMA), a common electron-beam photoresist. The possible applications of this type of polymer would be for stabilizing the carrier concentration, photochemical gating, or to create p-n junctions.[22,24,25] Figure 3a shows transport characteristics of device S5D2 before applying PMMA, with n-type carrier density of $7.4 \times 10^{11}$ cm$^{-2}$. Pronounced SdH oscillations corresponding to filling factor indices $\nu$=2, 6, 10, and 14 are clearly observed with the well-developed $\nu$=6 plateau occurring for $R_{xy}$ between 4 T and 6 T. Figure 3b shows results after the sample was coated with a PMMA layer and cured; the carrier density increased to $n = 2.5 \times 10^{12}$ cm$^{-2}$. Finally, the polymer layer on S5 was removed, and the results in Figure 3c show that the carrier density once again increased to $5.7 \times 10^{12}$ cm$^{-2}$. Increasing carrier density may be due to displacement or removal of the p-type doping agent from the EG surface and polymer residue (see Figure S2).

Our new protective precious-metal masking process for fabrication of graphene-based devices results in very low carrier concentrations and high mobility. Elsewhere, enhanced p-type doping in EG has been produced by gold adatoms but only after post-annealing to at least 700 °C.[26] Intercalation of hydrogen, gold, oxygen and other atoms beneath the EG buffer layer also can produce a p-type carrier concentration,[16] but these processes only occur at high temperatures. The temperatures encountered in our fabrication are never higher than 180 °C. Energy-dispersive X-ray spectroscopy (EDS) spectra showed only carbon, silicon, and trace atomic concentrations of oxygen. No indication of either Au or Pd was found on the aqua-regia-exposed EG regions. Thus, we believe that our results unequivocally show that extrinsic molecular doping is responsible for the low carrier concentration, and this doping process is initiated by aqua regia.

Both of the components of aqua regia, nitric acid and hydrochloric acid, are potent p-doping agents of graphene.[27,28] Since oxygen is the only significant component in the EDS spectra from as-fabricated graphene devices other than carbon and silicon, evidence points to $HNO_3$ as the primary doping agent, either by itself or by some chemical process in concert with other molecules present in the air. The observed changes in the level of carrier activity after exposure to the laboratory air support the notion of additional doping processes.

In summary, we have fabricated devices without polymer residues using epitaxial graphene and found that the carrier density is typically below $3\times10^{11}$ cm$^{-2}$ and is uniform on the same sample to within $1\times10^{11}$ cm$^{-2}$ after molecular doping by DAR. On sample S4 we observed highly quantized $\nu$=2 QHR plateaus with $R_{xy} \approx h/(2e^2)$ and near-zero longitudinal resistivity. Device S4D4 displays almost fully-quantized Hall plateaus, showing that EG is not significantly damaged by the deposition of precious metals (Pd, Au) or by diluted aqua regia, even after several immersion steps.

This new fabrication and doping process avoids organic contamination of devices based on epitaxial graphene grown on SiC. The method thus provides an alternate route for producing large-scale, highly ordered, low carrier-density epitaxial graphene for QHR standards and for the study of low-carrier density graphene.

**Acknowledgements**

The authors gratefully acknowledge support for Y.Y's work at NIST, federal grant #70NANB12H185.

# Supporting Information

## Low Carrier Density Epitaxial Graphene Devices On SiC

*Yanfei Yang*, Lung-I Huang, Yasuhiro Fukuyama, Fan-Hung Liu, Mariano A. Real, Paola Barbara, Chi-Te Liang, David B. Newell, and Randolph E. Elmquist*

### Surface characterization by AFM

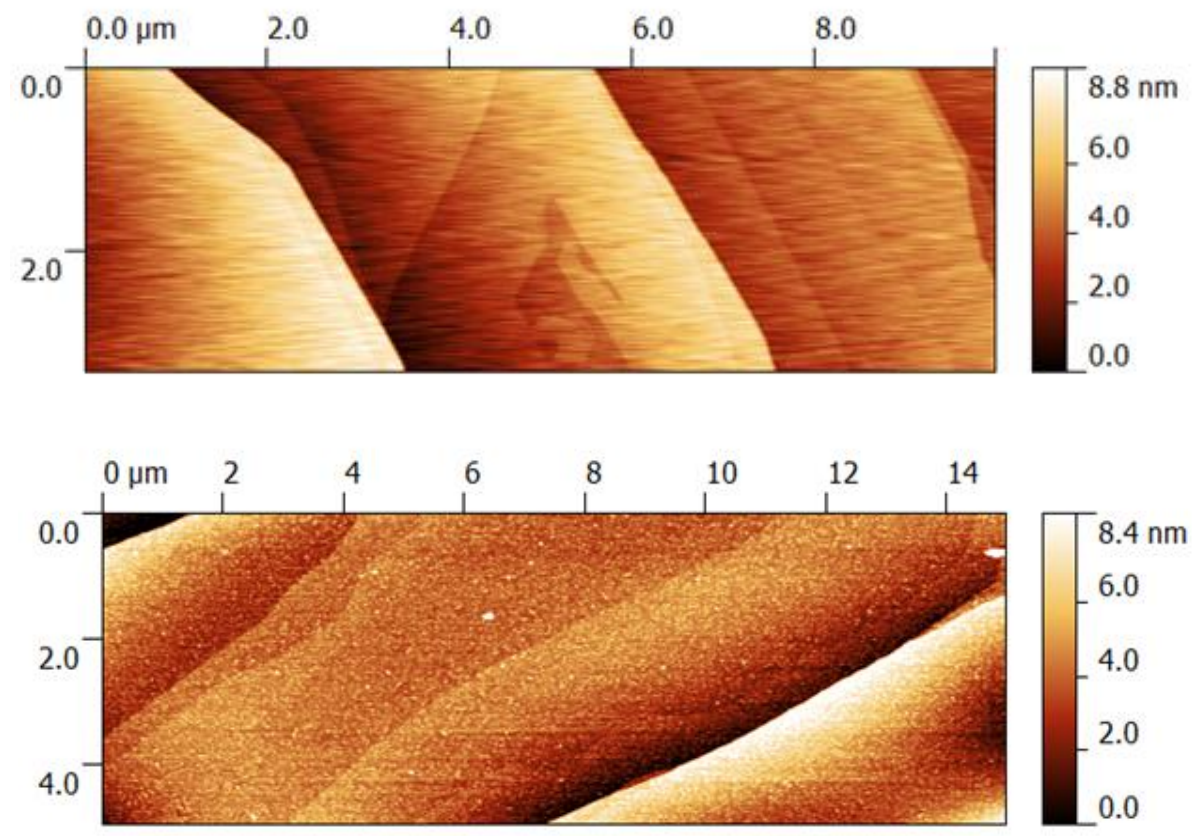

**Figure S1.** AFM tapping mode images (with height shown in scale bars) for sample S1 before (top) and after (bottom) device fabrication.

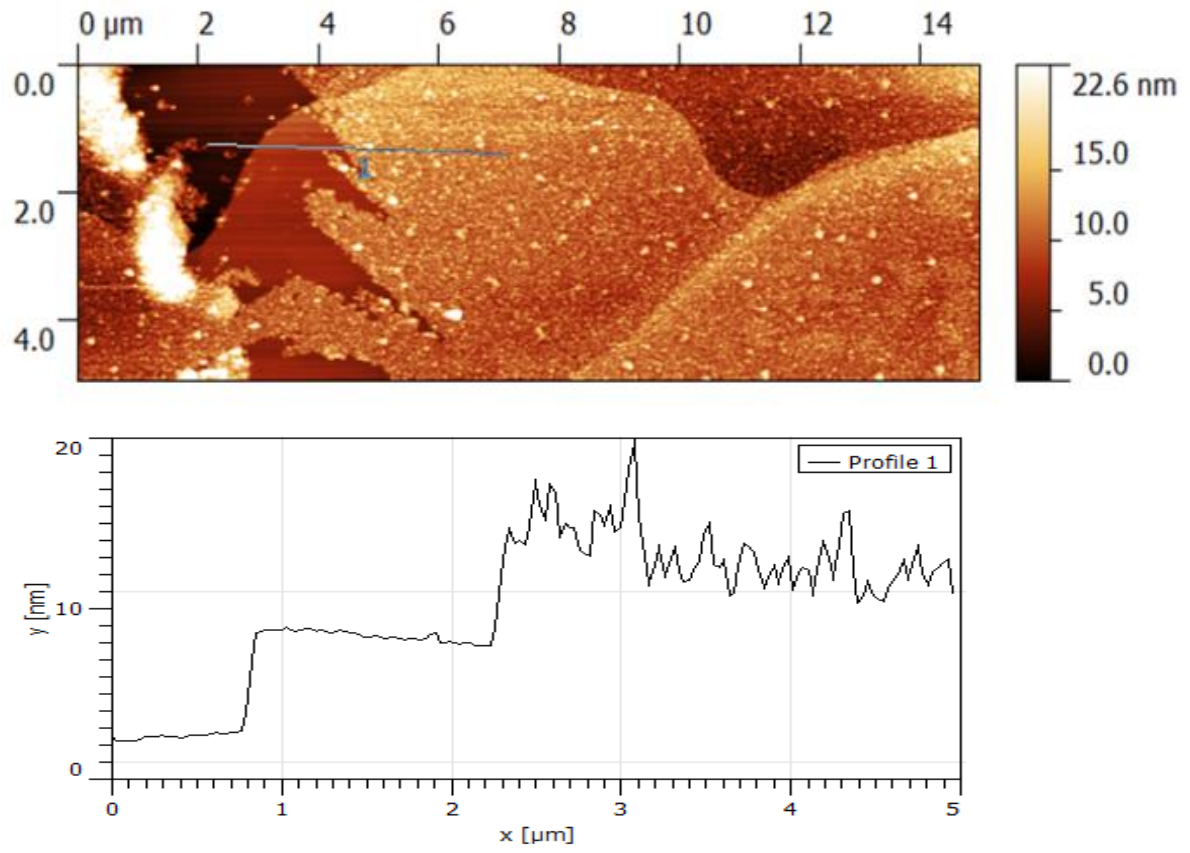

**Figure S2.** Top: AFM height image of sample S5 after removal of PMMA. To the left side, raised white areas may be polymer residue caused by bulk removal of PMMA from the adjacent EG surface. Bottom: Surface height contour produced along line (1) in the top image.

### Epitaxial graphene

Our graphene was grown on the Si-face of 6H-SiC(0001) substrates obtained from AGP Ltd. or 4H-SiC(0001) obtained from Cree, Inc.† All wafers were chemically-mechanically polished to produce nearly atomically-flat Si-face surfaces. The SiC wafers were diced into either square (7.6 mm x 7.6 mm) or rectangular (7.6 mm x 3.8 mm) samples. All samples originated from three SiC wafers. Samples 1, 2 and 3 were diced from the same wafer. Samples 1 and 2 were produced under identical conditions, in the same furnace run. Before fabrication, samples 1, 2, and 3 were kept in laboratory air for about two years after the EG growth. Sample 4 was produced four weeks before device fabrication and sample 5 was grown eight months prior to device fabrication.

**Fabrication Method**

*Step-1*: A metal layer (light green layer in Figure 1b) of thickness 15 nm to 30 nm is evaporated over the graphene sample surface after 10 minutes dehydration baking at 115 °C on a hotplate. This metal layer protects the graphene in the device area, including the Hall bar area and that underneath contact electrodes, from any residue contamination by photoresist or photoresist remover used in the rest of the fabrication process.

*Step 2*: A bi-layer photoresist of LOR3A/S1813 is spun on top of the metal blanket deposited in step-1 (Figure 1c). The sample is baked at 170 °C after the coating of LOR3A and at 115 °C after the coating of S1813 subsequently. Using an optical mask with a pattern of 4x4 arrays of Hall bars, we expose the Hall bar area by the i-line UV light with a dosage of ~100 mJ (Figure 1d). An undercut photoresist profile is achieved after developing the sample in CD-26 for 1 minute. Then a 50 nm thick layer of Au is evaporated on the sample (dark green layer in Figure 1e), followed by lift-off in PG-remover.

*Step-3*: A time-controlled Ar-RIE process (Figure1f) is applied to the whole sample to remove the first metal layer outside of the Hall bar area as well as the graphene underneath. As the metal covering the Hall bar area is much thicker, it will only be partially etched by the Ar plasma. Therefore, we obtain well patterned Hall bars covered by a metal mask after the Ar-RIE process (Figure 1g).

*Step-4:* Repeating step-2 but using the second optical mask, where electrodes are patterned to make contact with graphene Hall bars (Figure 1h) through the metal mask.

*Step-5:* A single photoresist layer of S1813 liquid is spun on the sample. An etching window in the Hall bar area is opened after UV exposure through a third optical mask and development in CD-26 (Figure 1i). Then the whole sample is soaked in fresh diluted aqua regia (by volume, $HNO_3$:HCl:$H_2O$ = 1:3:4) for ~45 seconds, which removes the metal mask that protects the graphene Hall bars. S1813 is resistant to DAR and may be stripped in acetone (Figure 1j).

**Device geometry**

We applied the above fabrication process to five different samples and obtained more than 40 devices with the same large Hall bar (0.6 mm x 0.1 mm) pattern as shown in Figure 1. Three 20 µm wide voltage probes (1,2,3,7,8,9) are located in the middle region of the Hall bar, 100 µm apart from each other, with graphene-metal contact area 3600 µm$^2$. Source and drain current contacts (5,10) have contact area 10000 µm$^2$.

**Device selection and characterization measurements**

Devices were selected by room-temperature four-terminal measurements of the sheet resistance $R_s$. All of the characterized devices had eight ohmic contacts at room temperature. Devices having $R_s \leq 21$ kΩ were selected for low-temperature transport characterization using lock-in amplifiers with a 13 Hz ac source-drain current at levels near 1 µA. All eight device contacts were wire-bonded to pins on TO-8 headers. The longitudinal voltage was recorded for contacts on both sides of the channel, using contact pairs (9,7) and (1,3), while the Hall voltage was measured using contact set (8,2) (see Figure 1a). The longitudinal resistivity, $\rho_{xx} = \frac{V_{xx}}{I_{xx}}\frac{L}{w}$, is the average of longitudinal voltage values on each side of the device divided by the current (measured simultaneously) and scaled by the aspect ratio of the channel (L is the spacing between voltage probes and w is the channel width). The yield, device location on the samples, and device-specific transport characteristics of all devices that were characterized are given below (Figure S3 and Table S1). Key to diagrams shown below:

Red squares: Devices characterized by magnetrotransport measurements and identified in table S1

Green squares: Uncharacterized devices with 8 usable contacts

Brown squares: Unusable devices

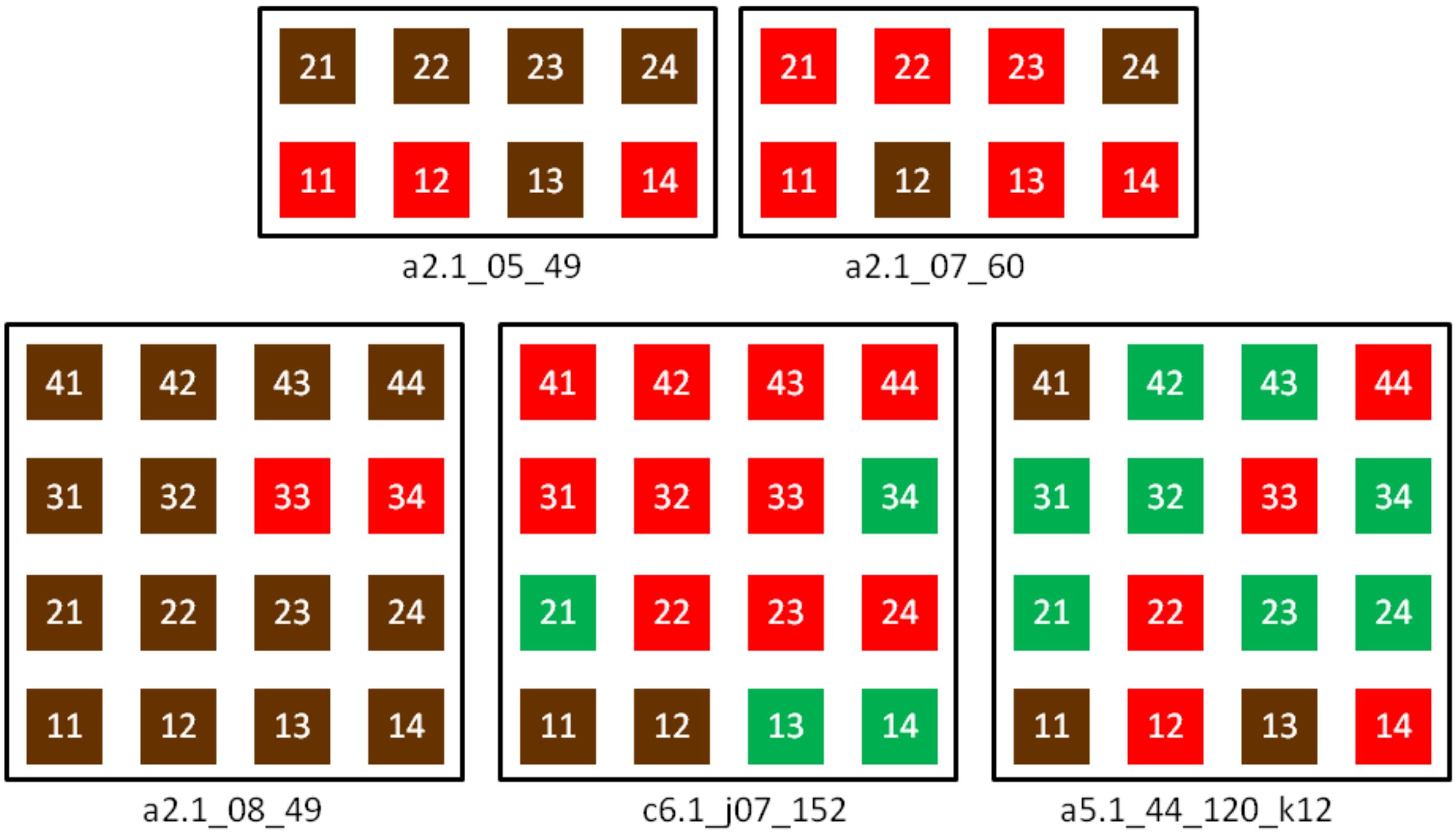

**Figure S3.** Diagram of devices on the five samples reported in the main text.

**Table S1.**Growth conditions and transport characteristics (for five samples reported in the main text) approximately one month after fabrication.

| Device | *T* (K) | Growth Conditions (°C/min) | Electron density *n* ($cm^{-2}$) | Mobility $\mu$ ($cm^2V^{-1}s^{-1}$) | $R_s$ (kΩ) *B*=0 | |
|---|---|---|---|---|---|---|
| a2.1_05_49_11 | 4.3 | 1900/30 | $1.3x10^{11}$ | 11000 | 4.4 | S1D1 |
| a2.1_05_49_12 | 4.3 | | $1.4x10^{11}$ | 6500 | 7.1 | S1D2 |
| a2.1_05_49_14 | 4.3 | | $5.1x10^{10}$ | 10200 | 11.9 | S1D3 |
| a2.1_08_49_33 | 1.65 | 1900/30 | $-8.5x10^{10}$ | 1060 | 69.2 | |
| a2.1_08_49_34 | 4.0 | | $4.2x10^{10}$ | 7400 | 20.2 | S2D1 |
| a2.1_07_60_14 | 4.3 | 1900/30 | $1.0\ x10^{10}$ | 1062 | 5.85 | |
| a2.1_07_60_11 | 4.3 | | $-5.4\ x10^{10}$ | -7911 | 14.6 | S3D1* |
| a2.1_07_60_13 | 4.3 | | $5.14\ x10^{10}$ | 10752 | 11.3 | S3D2 |
| a2.1_07_60_21 | 4.3 | | $3.98\ x10^{10}$ | 8581 | 18. 2 | S3D3 |
| a2.1_07_60_22 | 4.3 | | $6.94\ x10^{10}$ | 4362 | 20.6 | S3D4 |
| a2.1_07_60_23 | 4.3 | | $6.44\ x10^{10}$ | 8507 | 11.4 | S3D5 |
| c6.1_J07_152_24 | 1.5 | 1900/18 | $1.90x10^{11}$ | 1974 | 16.6 | |
| c6.1_J07_152_32 | 1.5 | | $2.01x10^{11}$ | 1928 | 16.1 | |
| c6.1_J07_152_42 | 1.5 | | $1.73x10^{11}$ | 4104 | 8.7 | |
| c6.1_J07_152_43 | 1.5 | | $2.57x10^{11}$ | 2096 | 11.6 | |
| c6.1_J07_152_44 | 1.5 | | $1.93x10^{11}$ | 5717 | 5.6 | |
| c6.1_J07_152_22 | 4.5 | | $2.04x10^{11}$ | 6649 | 4.6 | S4D1 |
| c6.1_J07_152_23 | 4.5 | | $1.91x10^{11}$ | 5966 | 5.5 | S4D2 |
| c6.1_J07_152_31 | 4.5 | | $3.11x10^{11}$ | 2571 | 7.8 | S4D3 |
| c6.1_J07_152_33 | 4.5 | | $2.35x10^{11}$ | 5195 | 5.1 | S4D4 |
| c6.1_J07_152_41 | 4.5 | | $3.24x10^{11}$ | 2128 | 9.0 | S4D5 |
| | | | | | | |
| a5.1_44_120_12 | 1.5 | 1950/15 | $8.03x10^{11}$ | 4974 | 3.1 | S5D1 |
| a5.1_44_120_14 | 1.5 | | $7.40x10^{11}$ | 4440 | 3.8 | S5D2 |
| a5.1_44_120_22 | 1.5 | | $8.24\ 10^{11}$ | 3961 | 3.8 | S5D3 |
| a5.1_44_120_33 | 1.5 | | $8.15\ 10^{11}$ | 3278 | 4.6 | S5D4 |
| a5.1_44_120_44 | 1.5 | | $6.29x10^{11}$ | 4357 | 4.5 | S5D5 |

**Device storage and post-processing**

We stored the devices in the laboratory, protected by stainless-steel caps designed for use with TO-8 headers. Device S4D4 was removed from the TO-8 header each time it was heat-treated or recycled using DAR. Device S5D2 was removed from the header for post-processing using PMMA.

**PMMA processing**

PMMA was spin-coated, dried at 170 °C, and removed using Microposit® remover 1165 and acetone in sequence at room temperature, followed by isopropyl alcohol rinsing.

[†]*Mention of commercial products or services does not imply endorsement by NIST, nor does it imply that such products or services are necessarily the best available for the purpose.*